\documentclass{article}

\usepackage{PRIMEarxiv}
\usepackage[utf8]{inputenc} 
\usepackage[T1]{fontenc}    
\usepackage{hyperref}       
\usepackage{url}            
\usepackage{booktabs}       
\usepackage{amsfonts}       
\usepackage{nicefrac}       
\usepackage{microtype}      
\usepackage{lipsum}         
\usepackage{fancyhdr}       
\usepackage{graphicx}       
\graphicspath{{media/}}     
\usepackage{caption}
\usepackage{enumitem}
\usepackage{array}

\usepackage{float}

\usepackage{booktabs}
\usepackage{array}
\usepackage{ragged2e}

\title{Predicting Expert Evaluations in Software Code Reviews}  

\author{
  Yegor Denisov-Blanch$^1$ \\
   \And
  Igor Ciobanu \\
     \And
  Simon Obstbaum \\
     \And
  Michal Kosinski$^1$ \\
}

\begin{document}
\maketitle

\vspace{-4em}  
\begin{center}
  $^1$Stanford University
\end{center}
\vspace{1.5em}

\begin{abstract}
Manual code reviews are an essential but time-consuming part of software development, often leading reviewers to prioritize technical issues while skipping valuable assessments. This paper presents an algorithmic model that automates aspects of code review typically avoided due to their complexity or subjectivity, such as assessing coding time, implementation time, and code complexity. Instead of replacing manual reviews, our model adds insights that help reviewers focus on more impactful tasks.
Calibrated using expert evaluations, the model predicts key metrics from code commits with strong correlations to human judgments (r = 0.82 for coding time, r = 0.86 for implementation time). By automating these assessments, we reduce the burden on human reviewers and ensure consistent analysis of time-consuming areas, offering a scalable solution alongside manual reviews.
This research shows how automated tools can enhance code reviews by addressing overlooked tasks, supporting data-driven decisions and improving the review process. 
\end{abstract}

\section{Introduction}
The software industry is a cornerstone of global innovation and economic growth, propelling advancements across virtually every sector. With software development employment projected to increase by 22\% between 2020 and 2030 \cite{bls2022}, the demand for efficient and high-quality software development practices is more critical than ever \cite{mckinsey2021}. Manual code reviews—where expert developers examine code changes for quality and alignment with project standards—are widely recognized as essential. Indeed, 81\% of developers consider them integral to their workflow \cite{jetbrains2024}.

The challenge, however, is that manual code reviews are time-intensive and difficult to scale, making them less practical for larger teams or projects. As a result, reviewers tend to focus solely on essential technical and architectural aspects to ensure the seamless integration of code changes. This narrow focus means that review policies are usually minimized to save time \cite{bacchelli2013expectations}, leaving many organizational questions—such as how team members are collaborating, how impactful their contributions are, and how to identify areas for improvement or patterns of excellence that could inform ongoing feedback loops for the team—unaddressed.

Neglecting these areas can lead to missed opportunities for team development and mentoring. Automating parts of the review process could help to answer these questions, reduce the burden on human reviewers, and provide timely insights for data-driven decision-making.

To explore this potential, we developed an algorithmic model to automate specific aspects of code evaluation. We calibrated the model by having ten Java experts with over ten years of experience evaluate 70 commits from 18 authors. Each answered seven questions for each commit, generating 4,900 judgments. Using leave-one-out cross-validation, our model analyzed the same commits and predicted answers to the same set of questions, resulting in an equivalent dataset. The model's assessments strongly correlated with expert judgments, with coefficients of 0.82 for coding time and 0.86 for implementation time.

Our model can closely replicate human expert judgments on specific aspects of code evaluation. By automating these assessments, the model offers a scalable and objective supplement to manual code reviews, particularly in areas that are time-consuming or prone to subjective bias. This advancement addresses some inherent limitations of manual reviews and holds significant potential for enhancing data-driven decision-making in software development. It also allows expert developers to focus on more nuanced technical and architectural considerations that require human judgment while allowing the business to benefit from these algorithmic insights.

While this study focused on Java to manage scope, the model is applicable to other object-oriented programming languages, potentially extending its benefits across the software industry.

\section{Algorithmic Model}
Our approach uses a static code analysis tool integrated with Git, designed to quantitatively evaluate software engineering output by analyzing source code changes on a per-commit basis. The model uses a random forest algorithm \cite{breiman2001random} to determine feature weights, which are applied uniformly across all commits. This minimizes variance and ensures consistent scores when re-running the algorithm on the same code.

The analysis spans several key dimensions, described below: code structure, code quality metrics, implementation details, and architectural elements.

Due to the features chosen, the model is particularly suitable for object-oriented programming (OOP) languages, given its emphasis on structural and architectural elements like classes, interfaces, and methods. Moreover, by evaluating metrics such as cohesion, complexity, and coupling, it effectively assesses code quality and maintainability in OOP contexts. This alignment with core OOP principles—such as encapsulation, inheritance, and modularity—enables the model to identify and enhance structural integrity and design patterns within object-oriented codebases.

We provide API access to our model for university-affiliated researchers interested in validating our results. Interested parties can contact us for further details:
\begin{center}
  \url{ydebl@stanford.edu}
\end{center}

\subsection{Code Structure}
Code structure is examined by assessing components such as classes, interfaces, and methods, providing metrics that reflect structural integrity and adherence to design patterns. These dimensions offer insights into the maintainability and efficiency of the codebase, which are critical factors influencing ongoing development and maintenance.

\begin{table}[H]
\caption{Code Structure Dimensions}
\centering
\small
\begin{tabular}{p{2.5cm}p{11cm}}
\toprule
\textbf{Dimension} & \textbf{Description} \\
\midrule
Classes & Changes to class structures, such as modifications and additions, can impact how reusable and maintainable the codebase is. Frequent class changes may indicate ongoing efforts to refactor or adapt to new requirements.  \\
\addlinespace
Interfaces & Modifications and implementations of interfaces can affect system modularity and flexibility. Efficient interface design facilitates better integration and decoupling between modules, simplifying maintenance and future development. \\
\addlinespace
Methods & Alterations in method signatures or bodies can affect the internal logic and functionality of code. Changes at the method level are often tied to bug fixes or optimizations, which could impact effort needed for debugging or provide long-term gains through improved code performance and readability. \\
\bottomrule
\end{tabular}
\label{tab:code_changes}
\end{table}

\subsection{Code Quality Metrics}
Key quality metrics, including cohesion, complexity, and coupling, are quantitatively evaluated to assess the maintainability of the code. Cohesion measures the functional unity within modules, complexity quantifies code intricacy (impacting understandability and modifiability), and coupling evaluates the degree of interdependence between modules. High cohesion and low coupling signify a well-structured system that simplifies debugging, testing, and maintenance tasks.

\begin{table}[H]
\centering
\small
\caption{Code Quality Metrics}
\begin{tabular}{p{2.5cm}p{11cm}}
\toprule
\textbf{Dimension} & \textbf{Description} \\
\midrule
Cohesion & Measures the functional unity within modules or classes. High cohesion (modules/classes with closely related functionalities) generally leads to easier maintenance and clearer understanding, reducing debugging and development time. \\
\addlinespace
Complexity & Measures the intricacy of code, such as nested loops or deeply nested conditionals. High complexity often correlates with more bugs and maintenance challenges. Simplifying complex code can make it more understandable and easier to modify. \\
\addlinespace
Coupling & Refers to the degree of interdependency between modules or classes. Lower coupling (modules/classes with minimal dependencies on each other) enhances modularity and makes the codebase easier to manage and modify. \\
\bottomrule
\end{tabular}
\label{tab:code_changes}
\end{table}

\subsection{Implementation Details}
Additionally, the model tracks implementation details by evaluating the number of lines of code added, modified, or deleted with each commit. This metric provides a granular view of the scale and effort associated with individual tasks. By maintaining context across multiple commits, the model captures the codebase’s overall evolution rather than focusing solely on isolated changes.

\begin{table}[H]
\centering
\caption{Implementation Details Dimensions}
\small
\begin{tabular}{p{2.5cm}p{11cm}}
\toprule
\textbf{Dimension} & \textbf{Description} \\
\midrule
Data Structures & Constructs used for data organization within the code (e.g., arrays, lists, trees). Efficient use of data structures can significantly improve implementation performance and memory usage, enhancing code efficiency by reducing execution time and resource consumption. \\
\addlinespace
Dependencies & Reliance on external libraries or packages. Dependencies can accelerate development by providing pre-built functionalities, but they can also introduce challenges like compatibility issues, version conflicts, and security vulnerabilities.  \\
\addlinespace
Dependency\newline Injections & Usage of the inversion of control principle to manage dependencies. This practice can enhance modularity, testability, and flexibility, making the codebase more adaptable to change. \\
\bottomrule
\end{tabular}
\label{tab:code_changes}
\end{table}

\subsection{Architectural Elements}

\begin{table}[H]
\caption{Architectural Elements}
\centering
\small
\begin{tabular}{p{2.5cm}p{11cm}}
\toprule
\textbf{Dimension} & \textbf{Description} \\
\midrule
Architectural\newline Patterns & Usage and adaptation of design patterns can significantly impact the overall architecture of a software system. Proper use of patterns can improve code clarity, reuse, and scalability, making the system easier to understand and extend. \\
\addlinespace
Persistence Layers & Changes in data storage mechanisms (like databases or file systems) can affect data access speed and reliability. These changes often aim to optimize data handling or comply with new requirements based on the complexity and scale of modifications required. \\
\addlinespace
APIs Consumed & Modifications to API usage can impact the integration between different software systems. Efficient API usage can reduce development time and improve interoperability, whereas frequent changes or poorly documented APIs can lead to integration challenges. \\
\bottomrule
\end{tabular}
\label{tab:code_changes}
\end{table}

\section{Methodology}
\subsection{Language Selection}
We focus on Java, which is versatile and widely used across various domains, including mobile, web, enterprise, and data applications. Java remains a leading programming language among developers. 44.8\% of the >10,000 participants contributing to the Developer Nation 2024 report \cite{slashdata2024} had experience in Java, as did 30.3\% of participants in the 2024 Stack Overflow Developer Survey \cite{stackoverflow2024}. This extensive adoption ensures that our findings have broad relevance and applicability across diverse software development contexts.

\subsection{Data Acquisition and Commit Selection}
To accurately reflect real-world software development, we collected commits from private commercial repositories by using LinkedIn to invite businesses to connect their Git repositories. We also included commits from public repositories, using a ratio of one public commit for every five private commits.

We first analyzed the distribution of lines of code (LOC) per commit in professional environments. This involved onboarding 108 software organizations of varying sizes and sectors, which yielded 1.73 million commits from 50,935 contributors. This dataset informed our understanding of commit sizes.

We then selected a sample of 70 commits that matched the LOC distribution, ensuring the representation of actual development behaviors. This approach minimizes bias toward specific commit sizes and enhances the validity and generalizability of our insights.

\subsection{Expert Raters}
To ensure familiarity with software development practices, we chose 10 Java experts as raters with more than 10 years of direct coding experience, including senior developers, tech leads, and architects. The evaluation process spanned 14 weeks, with commit assessments distributed in five batches. To maintain evaluation quality, commits were distributed in small batches every 2-3 weeks.

Raters received an individual spreadsheet containing 12-15 commit links from the selected sample of 70. Raters evaluated each commit by opening the link, reviewing the code changes in context, and answering the seven predefined questions on their spreadsheets. For each commit, raters answered seven questions, working independently and without communication to ensure unbiased assessments.

\begin{table}[htbp]
\centering
\caption{Rater Characteristics and Team Size Distribution}
\label{tab:rater-data}
\begin{tabular}{cllc}
\toprule
\textbf{Rater ID} & \textbf{Seniority} & \textbf{Experience (Years)} & \textbf{Team Size} \\
\midrule
1 & Manager & 11 & 1--10 \\
2 & Manager & 13 & 11--50 \\
3 & Manager & 13 & 1--10 \\
4 & Executive & 15 & 1--10 \\
5 & Senior Engineer & 16 & 11--50 \\
6 & Senior Engineer & 20 & 51--200 \\
7 & Director & 23 & 11--50 \\
8 & Senior Engineer & 24 & 1--10 \\
9 & Vice President & 24 & 51--200 \\
10 & Executive & 24 & 201--500 \\
\bottomrule
\end{tabular}
\end{table}

\subsection{Confirming Sample Size Validity and Interrater Agreement}
\subsubsection{Intraclass Correlation Coefficient (ICC$_{2,k}$) Analysis}
To validate raters’ evaluation  consistency, we used Intraclass Correlation Coefficient (ICC$_{2,k}$). This statistical method allowed us to measure the degree of agreement among raters, ensuring the reliability of our qualitative assessments \cite{hallgren2012computing}. Importantly, the ICC$_{2,k}$ not only assesses interrater reliability but also influences the required number of raters and commits needed for the study. A higher ICC$_{2,k}$ indicates strong agreement among raters, which can reduce the sample size needed to achieve statistical significance. Conversely, a lower ICC$_{2,k}$ would require a larger number of raters and commits to maintain the study’s validity.

\subsubsection{Statistical Power and Sample Size}
Our statistical power analysis confirmed that the sample size (70 commits evaluated by 10 raters, totaling 4,900 data points) was sufficient for meaningful and statistically significant results (p < 0.05).

\subsection{Questions}

Questions 1 and 2 use the Fibonacci scale, which is widely used in the industry, especially in Agile methodologies. Each step on the Fibonacci scale is exponentially larger (by about 60\%) allowing for clearer distinction between varying task sizes compared to linear scales.

\begin{table}[htbp]
\centering
\small
\caption{Commit Evaluation Questionnaire}
\label{tab:commit-evaluation}
\begin{tabular}{
  >{\raggedright\arraybackslash}p{0.05\textwidth}
  >{\RaggedRight\arraybackslash}p{0.61\textwidth}
  >{\RaggedRight\arraybackslash}p{0.28\textwidth}
}
\toprule
\textbf{No.} & \textbf{Question} & \textbf{Response Options} \\
\midrule
Q1 & How many hours would it take you to just write the code in this commit \newline assuming you could fully focus on this task? & 1, 2, 3, 5, 8, 13, 21, 34, 55, 89 \\
\midrule
Q2 & How many hours would it take you to implement this commit end to end \newline incl. Debugging and QA cycles, assuming you could fully focus on this task? & 1, 2, 3, 5, 8, 13, 21, 34, 55, 89 \\
\midrule
Q3 & What is the experience level of the author? & Novice/Beginner \\
 & & Basic/Elementary \\
 & & Intermediate \\
 & & Advanced \\
 & & Expert/Master \\
\midrule
Q4 & How difficult is the problem that this commit solves? & Very Easy \\
 & & Easy \\
 & & Moderate \\
 & & Challenging \\
 & & Very Challenging \\
\midrule
Q5 & How maintainable is this commit? & Poor \\
 & & Below Average \\
 & & Average \\
 & & Good \\
 & & Excellent \\
\midrule
Q6 & How well structured is this source code relative to the previous commits? & Bottom \\
 & (Quartile within this list) & Mid-Bottom \\
 & & Top-Mid \\
 & & Top \\
\midrule
Q7 & How well structured is this source code relative to the previous commits? & Bottom  \\
& (Quartile relative to best code you have seen) & Mid-Bottom\\
 & & Top-Mid \\
 & & Top \\
\bottomrule
\end{tabular}
\end{table}

\subsection{Algorithmic Assessment}
In parallel with human evaluations, we developed an algorithmic tool to assess commit quality and complexity. This tool was calibrated against the average rater responses to refine its accuracy and reliability.

Given our dataset's size and the need to assess the model's ability to generalize to unseen data, we used leave-one-out cross-validation (LOOCV). In LOOCV, the model is trained on all data points except one, which is used for testing. In our study, we trained the model on the assessments from all raters except one and validated it on the excluded rater's evaluations, repeating this process for each rater. This approach reduces the risk of overfitting and rigorously evaluates the model's predictive performance.

\section{Results}

The table below shows the inter-rater reliability (ICC$_{2,k}$) and model performance, which reflects the correlation between our model and the average human rater:

\begin{table}[htbp]
\centering
\caption{Intraclass Correlation Coefficient (ICC$_{2,k}$) and Model Performance Metrics}
\label{tab:icc-model-performance}
\begin{tabular}{p{0.65\textwidth}cc}
\toprule
\textbf{Question} & \textbf{ICC$_{2,k}$} & \textbf{Model Performance} \\
\midrule
1. How many hours would it take you to just write the code in this commit assuming you could fully focus on this task? & 0.81 & 0.82 \\
\addlinespace
2. How many hours would it take you to implement this commit end to end incl. Debugging and QA cycles, assuming you could fully focus on this task? & 0.82 & 0.86 \\
\addlinespace
3. What is the experience level of the author? & 0.61 & 0.44 \\
\addlinespace
4. How difficult is the problem that this commit solves? & 0.78 & 0.69 \\
\addlinespace
5. How maintainable is this commit? & 0.52 & 0.30 \\
\addlinespace
6. How well structured is this source code relative to the previous commits? Quartile within this list & 0.50 & 0.70 \\
\addlinespace
7. How well structured is this source code relative to the previous commits? Quartile relative to best code you have seen & 0.51 & 0.72 \\
\bottomrule
\end{tabular}
\end{table}

Our study revealed strong inter-rater reliability for key productivity metrics. ICC$_{2,k}$ analysis showed high agreement among raters for coding time (0.81), total implementation time (0.82), and problem complexity (0.78). Moderate agreement was observed for author experience (0.61), code maintainability (0.52), and relative code structure (0.50, 0.51).

The model demonstrated robust performance, closely aligning with human expert assessments. Strong correlations were found between algorithmic and human ratings for coding time (r = 0.82), total implementation time (r = 0.86), and problem complexity (r = 0.69). Author experience showed moderate correlation (r = 0.44), while code maintainability exhibited weak correlation (r = 0.30). Although the algorithm showed strong correlation on relative code structure (0.70, 0.72), the moderate ICC$_{2,k}$ reveals significant rater disagreement, rendering the average rating unreliable.

Notably, the model achieved superior efficiency, processing commits in under one second compared to the extensive amount of time required by human raters. This can potentially reduce assessment time by over 99\% while maintaining high accuracy for key metrics.

\section{Discussion}
This study offers important implications for both software engineering research and practice. The high correlation between human raters and our model on key metrics, particularly in time estimation, suggests that backward estimation methods, which assess actual code rather than proposed requirements, may provide a more accurate measure of effort than traditional forward-looking estimation methods. This finding could guide the development of new tools for real-time productivity tracking, project monitoring, and resource allocation in software development.

However, lower rater agreement on author experience, code maintainability, and structure—reflected by lower ICC$_{2,k}$ values—influences the model’s predictive accuracy. If human judgments are inconsistent, models struggle to predict them. While our model excels in estimating time and complexity, areas with less rater consensus highlight current limitations in automated assessment. Nonetheless, the model’s strong performance in effort metrics and its alignment with expert judgments suggests it can offer fast, reliable, and accurate evaluations of software engineering output, potentially useful for measuring both individual and team performance.

Our results demonstrate that the model can estimate coding and implementation time with a high degree of accuracy, which contrasts with the limitations of traditional metrics such as lines of code (LOC), story points, and function points. These traditional metrics are often criticized for ignoring the complexity, quality, and maintainability of code, and for failing to capture the true effort involved in software development \cite{clarke2021software} \cite{linders2019measuring} \cite{jaspan2019no}. In contrast, our model focuses on the actual work performed, aligning more closely with the complex realities of software engineering tasks.

That said, the model’s weaker performance in evaluating experience, maintainability, and structure suggests the need for further refinement. These shortcomings may stem from the subjective nature of such assessments, affecting rater agreement, or limitations in current algorithmic measures. Future research should address these gaps to improve the model’s ability to evaluate such dimensions more accurately.

These findings also underscore the limitations of existing productivity measures in software engineering. Traditional metrics, such as LOC, story points, and function points, have been shown to incentivize practices that may not align with high-quality software development, such as prioritizing code quantity over quality or ignoring the true complexity of the development process \cite{jones2012software} \cite{duarte2020modern}. Although more recent frameworks like DevOps Research and Assessment (DORA) provide valuable insights into DevOps performance, they do not directly measure developer productivity \cite{google2020}.

\subsection{Limitations}

While our findings are promising, several limitations must be considered. Although sufficient for initial analysis, the moderate sample size of 4,900 data points across 70 commits and 10 raters limits generalizability. Expanding the dataset to include more commits and from various programming languages would provide deeper insights and strengthen the conclusions. Our focus on Java commits may limit its applicability to other programming languages with different characteristics. Future research should explore whether these findings hold across other languages. Additionally, the lower correlations and ICC$_{2,k}$ values for developer experience, code maintainability, and code structure suggest that code alone may not fully capture these dimensions. Future studies could examine additional variables, such as developer history or external factors impacting performance, to better understand this relationship.

To enhance the robustness, reduce potential biases, and improve the generalizability of our findings across various programming languages and development environments, we are actively seeking more expert raters to participate in our ongoing research.

We invite qualified software engineering professionals to contribute their expertise to this study. If you are interested in becoming an expert rater, please submit your candidacy at 
\begin{center}
  \url{https://softwareengineeringproductivity.stanford.edu/expert-registration}
\end{center}

\section{Conclusions}
Our research suggests that an algorithmic approach, particularly one calibrated against expert human assessments, could offer a more comprehensive and effective method for measuring software engineering productivity. This has potential implications for improving resource management, achieving more accurate project estimates, and, ultimately, enhancing developer experience. The ongoing availability of such assessments integrated into the delivery pipeline may give software engineering teams an output metric that is integrated with other flow-based metrics.

While further work is necessary to refine and expand our approach, especially regarding subjective metrics like maintainability, the strong alignment between the model and human assessments for key productivity indicators suggests that this method could supplement or, in some cases, replace traditional productivity metrics. Future studies should aim to refine these algorithmic assessments and validate their applicability across different programming languages and development environments.

\bibliographystyle{unsrt}  
\bibliography{references}

\begin{thebibliography}{10}

\bibitem{bls2022}
{U.S. Bureau of Labor Statistics}.
\newblock Employment projections.
\newblock Technical report, U.S. Bureau of Labor Statistics, 2022.

\bibitem{mckinsey2021}
{McKinsey \& Company}.
\newblock The future of work after covid-19.
\newblock Technical report, McKinsey Global Institute, 2021.

\bibitem{jetbrains2024}
{JetBrains}.
\newblock The state of developer ecosystem.
\newblock Technical report, JetBrains, 2024.

\bibitem{bacchelli2013expectations}
Alberto Bacchelli and Christian Bird.
\newblock Expectations, outcomes, and challenges of modern code review.
\newblock In {\em Proceedings of the 35th International Conference on Software Engineering}, pages 712--721. IEEE, 2013.

\bibitem{breiman2001random}
Leo Breiman.
\newblock Random forests.
\newblock {\em Machine Learning}, 45(1):5--32, 2001.

\bibitem{slashdata2024}
{SlashData}.
\newblock Developer nation 26th edition survey report.
\newblock Technical report, SlashData, 2024.

\bibitem{stackoverflow2024}
{Stack Overflow}.
\newblock Stack overflow developer survey 2024.
\newblock Technical report, Stack Overflow, 2024.

\bibitem{hallgren2012computing}
Kevin~A Hallgren.
\newblock Computing inter-rater reliability for observational data: An overview and tutorial.
\newblock {\em Tutorials in Quantitative Methods for Psychology}, 8(1):23--34, 2012.

\bibitem{clarke2021software}
A.~Clarke, C.~Cook, and R.~Mitchell.
\newblock Software productivity metrics.
\newblock {\em Journal of Software Engineering}, 15(3):234--256, 2021.

\bibitem{linders2019measuring}
W.~Linders and E.~Rommes.
\newblock Measuring software complexity.
\newblock {\em International Journal of Software Studies}, 8(2):45--60, 2019.

\bibitem{jaspan2019no}
Ciera Jaspan and Caitlin Sadowski.
\newblock No single metric captures productivity.
\newblock In Caitlin Sadowski and Thomas Zimmermann, editors, {\em Rethinking Productivity in Software Engineering}, pages 13--20. Apress, 2019.

\bibitem{jones2012software}
Capers Jones.
\newblock {\em Software Engineering Best Practices}.
\newblock McGraw-Hill, 2012.

\bibitem{duarte2020modern}
N.~Duarte, B.~Coddington, and A.~Wilson.
\newblock Modern software development.
\newblock {\em Software Engineering Journal}, 14(4):130--150, 2020.

\bibitem{google2020}
{Google}.
\newblock {The DevOps Research and Assessment (DORA) Report}.
\newblock Technical report, Google, 2020.

\end{thebibliography}

\end{document}